\documentclass[twocolumn,prl,showpacs,preprintnumbers,amsmath,amssymb]{revtex4}
\usepackage{graphicx}
\usepackage{dcolumn}
\usepackage{bm}
\usepackage{tabularx}
\usepackage{multirow}

\begin{document}

\title {Rock-salt SnS and SnSe: Native Topological Crystalline Insulators}

\author{Yan Sun$^{1,2,3}$}
\author{Zhicheng Zhong$^{4}$}
\author{Tomonori Shirakawa$^{2,5,6}$}
\author{Cesare Franchini$^3$}
\author{Dianzhong Li$^1$}
\author{Yiyi Li$^1$}
\author{Seiji Yunoki$^{2,5,6,7}$}
\author{Xing-Qiu Chen$^1$}
\email[Corresponding author:]{xingqiu.chen@imr.ac.cn}

\affiliation{$^1$ Shenyang National Laboratory for Materials
Science, Institute of Metal Research, Chinese Academy of Sciences,
Shenyang 110016, China,}

\affiliation{$^2$Computational Condensed Matter Physics Laboratory,
RIKEN, Wako, Saitama 351-0198, Japan,}

\affiliation{$^3$Computational Materials Physics, Faculty of
Physics, University of Vienna, Sensengasse 8/7, A-1090 Vienna,
Austria,}

\affiliation{$^{4}$Institute of Solid State Physics, Vienna
University of Technology, 1040 Vienna, Austria, }

\affiliation{$^5$CREST, Japan Science and Technology Agency (JST),
Kawaguchi, Saitama 332-0012, Japan, }

\affiliation{$^6$Computational Materials Science Research Team,
RIKEN Advanced Institute for Computational Science (AICS), Kobe,
Hyogo 650-0047, Japan,}

\affiliation{$^7$Computational Quantum Matter Reasearch Team, RIKEN
Center for Emergent Matter Science (CEMS), Wako, Saitama 351-0198,
Japan}

\date{\today}

\begin{abstract}

Unlike time-reversal topological insulators, surface metallic states with Dirac cone dispersion
in the recently discovered topological crystalline insulators (TCIs) are protected by crystal symmetry.
To date, TCI behaviors have been observed in SnTe and the related alloys Pb$_{1-x}$Sn$_{x}$Se/Te,
which incorporate heavy elements with large spin-orbit coupling (SOC). Here, by combining
first-principles and {\it ab initio} tight-binding calculations, we report the formation
of a TCI in the relatively lighter rock-salt SnS and SnSe. This TCI is characterized
by an even number of Dirac cones at the high-symmetry (001), (110) and (111) surfaces,
which are protected by the reflection symmetry with respect to the ($\overline{1}$10) mirror plane.
We find that both SnS and SnSe have an intrinsically inverted band structure and the SOC is necessary
only to open the bulk band gap. The bulk band gap evolution upon volume expansion reveals
a topological transition from an ambient pressure TCI to a topologically trivial insulator.
Our results indicate that the SOC alone is not sufficient to drive the topological transition.
\end{abstract}


\maketitle

Since the discovery of Z$_2$ topological insulators (TIs)~\cite{hasan,xiao},
band topological properties
in condensed matter physics have attracted increasing interest as a
new physical paradigm, which also shows great promise for
potentially revolutionary applications in quantum computing and
spintronics. TIs possess a non-trivial time-reversal Z$_2$ topological invariant
and the topological characteristics are manifested by the presence
of an odd number of linearly dispersing Dirac cones at the crystal surfaces.
These surface metallic states are due to large
spin-orbit coupling (SOC) and are protected by time-reversal
symmetry~\cite{hasan,xiao}.

In 2011, Liang Fu proposed a theoretical model for an alternative
class of topological states, named topological crystalline
insulators (TCIs), in which the gapless surface states are protected
not by time-reversal symmetry but by crystal
symmetry~\cite{fu,zaanen}.
Up to now, the only reported TCIs are
the narrow band gap semiconductor SnTe and the related alloys
Pb$_{1-x}$Sn$_{x}$Se/Te~\cite{fu2,dziawa,tanaka,Xu,fu3}. Very
recently, Barone {\em et al.} have theoretically predicted that a
suitable combination of applied pressure and alloying can turn
rock-salt lead chalcogenides, such as PbSe, PbTe, and PbS, into
TCIs~\cite{barone}.
The most prominent feature of this class of TCIs is the presence of
an even, not odd as in TIs, number of Dirac cones which lay on surface
terminations oriented perpendicular to the mirror symmetry planes.
It is shown that the necessary conditions for the band
inversions to occur in all these TCIs are (i) a strongly asymmetric hybridization
between cation (anion) \emph{s} and anion (cation) \emph{p} states
and (ii) a sizable SOC strength, similarly in time-reversal
TIs ~\cite{xia,zhang,lu,james,xiao2,sun,sun2,chen2}. Large SOC is
recognized to be a crucial ingredient to form possible TCIs also in
pyrochlore oxides $A_2$Ir$_2$O$_7$, where $A$ is a rare-earth element~\cite{Kargarian13}.
However,
TCIs can be considered as the counterpart of
TIs in materials without SOC~\cite{fu}. Thus, it
is of fundamental importance to seek a manifestation of the
non-trivial crystalline topology in materials composed of
constituents with lighter mass and thus smaller SOC, for which
the SOC effect is detached from the formation of TCIs.

In this Letter, through first-principles calculations along with
Wannier functions based {\it ab initio} tight-binding (TB) modeling, we
report that rock-salt SnS and SnSe are both TCIs in their native phase without any
alloying or applied strain/pressure. We find that their inverted band order
is induced by chemical bonding and crystal field, whereas the SOC effect is
only to open the bulk band gap. This non-trivial topological state is
substantiated by the emergence of an even number of Dirac cones
at the high-symmetry crystal surfaces perpendicular to the ($\overline{1}$10) mirror
symmetry plane.
We also demonstrate that a topological transition occurs to a trivial insulator
upon volume expansion.

First-principles calculations based on density functional theory
(DFT) are performed in the generalized gradient approximation (GGA),
following the Perdew-Burke-Ernzerhof parametrization
scheme~\cite{pbe}, with the projected augmented wave method as
implemented in the Vienna \emph{Ab initio} Simulation Package ({\small
VASP})~\cite{kresse,kresse2}. The energy cutoff is set to be 500 eV.
The TB matrix elements are calculated by projection onto
maximally localized Wannier orbitals~\cite{marzari,souza,mostofi},
using the VASP2WANNIER90 interface~\cite{Franchini}.

Early experimental characterizations~\cite{mariano,Nikolic,Palatnik}
have found that both SnS and SnSe crystallize in low temperatures
\emph{Pnma} GeS-type orthorhombic phase~\cite{Chattopadhyay} and
that at high temperatures two metastable orthorhombic \emph{Cmcm}
TlI-type~\cite{Chattopadhyay} and rock-salt cubic NaCl-type phases~\cite{Palatnik}
(Fig. 1) exist. Here, we focus on the
rock-salt structure, which has been shown to be stable under
epitaxial growth of SnSe and SnS on a NaCl substrate with lattice
constants  $\rm a_{Expt}^{SnSe}$ = 5.99 \AA\,~\cite{mariano} and
$\rm a_{Expt}^{SnS}$ = 6.00 \AA\,~\cite{mariano} and
5.80\AA\,~\cite{Bilenkii}, respectively. Our first-principles calculations find that
the optimized lattice constants are $\rm a_{Theo}^{SnSe}$ = 6.05 \AA\,
and $\rm a_{Theo}^{SnS}$ = 5.85 \AA\, for SnSe and SnS, respectively,
in good agreement with the experimental values.

\begin{figure}[htbp]
\begin{center}
\includegraphics[width=0.45\textwidth]{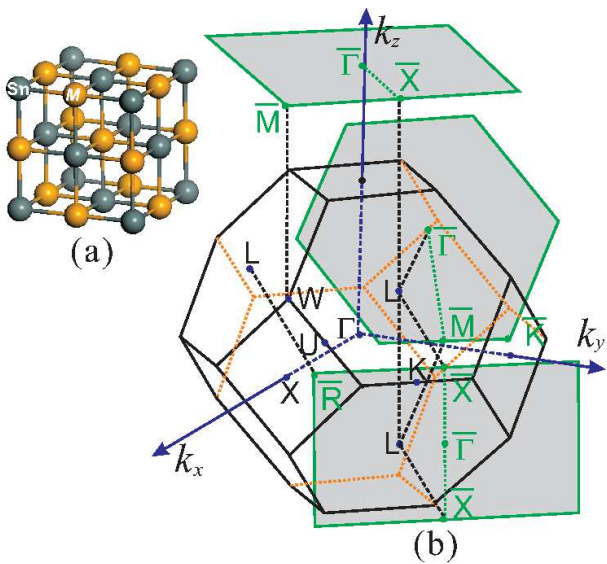}
\end{center}
\caption{(color online) (a) Sn$M$ ($M$ = S and Se) rock-salt lattice
structure and (b) face cubic centered (FCC) Brillouin zone (BZ). Two
dimensional BZ projected onto the (001), (110), and (111) surfaces
are also shown in (b). These three surfaces are all perpendicular to
the ($\overline{1}$10) mirror symmetry plane. } \label{crystal}
\end{figure}

\begin{figure}[htbp]
\begin{center}
\includegraphics[width=0.45\textwidth]{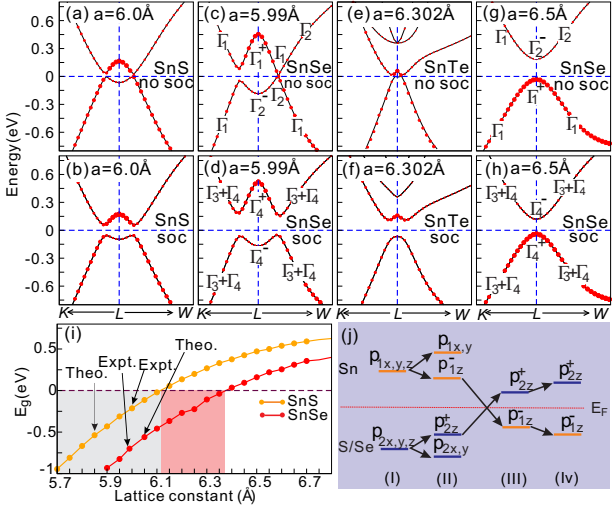}
\end{center}
\caption{(color online) (a-h): Electronic band structures obtained
by first-principles calculations for SnS (a,b), SnSe (c,d,g,h), and
SnTe (e,f) with and without SOC. The size of red dots in (a)-(h)
is proportional to the amount of contributing weight in each band
from S (a,b), Se (c,d,g,h), or Te (e,f) atoms. In (a)-(f), the experimental lattice
constant $a$ (indicated in the figures) is used~\cite{a_SnTe},
whereas in (g) and (h) an expanded lattice constant is used.
The irreducible representations of the bands closest to Fermi energy (set to be 0)
are also indicated~\cite{note3}.
(i): The evolution of the band gap $E_g$ calculated with SOC as a
function of the lattice constants for SnS and SnSe. The energy
gap $E_g$ is defined as the energy difference between the top of the
valence band and the bottom of the conduction band at $L$,
i.e., $E_g=E^{L}_{-} - E^{L}_{+}$, where $E^L_p$ is the
band energy with parity $p\,(=+,-)$ at $L$.
Thus, $E_g$ is negative when the bands are in inverted order.
The optimized and experimental
lattice constants are indicated by arrows, all located within the
non-trivial topological region. (j) A schematic energy level diagram
around $L$: (I) atomic limit,
(II) including hybridization with a large lattice constant,
(III) with the experimental lattice constant, and
(IV) inclusion of SOC.
The band is inverted already in (III) without SOC. Here
$E_F$ stands for Fermi energy. The signs $(+,-)$ denote
the parities of the corresponding $p$-like orbitals.
}\label{band_inv}
\end{figure}

{\em Bulk band structure.} The calculated bulk band structures along
the high symmetry lines around $L$ in the Brillouin zone (BZ) are shown in
Fig.~\ref{band_inv} (a)--(d), and for comparison the results for those
of the isostructural TCI SnTe are also plotted in
Fig.~\ref{band_inv} (e) and (f)~\cite{fu2,tanaka}.
The band structures of SnS, SnSe, and
SnTe display similar features, which are summarized as follows:
(\emph{i}) Without SOC, all compounds have a gapless three
dimensional (3D) Dirac cone located in the vicinity of the
high symmetry $L$ point along the $L$-$W$ line;
%in the Brillouin zone (BZ);
(\emph{ii}) Level anticrossing occurs once SOC is included
and the 3D Dirac cone is broken with opening a finite band gap;
(\emph{iii}) The top of the valence band
and the bottom of the conduction band at and near $L$ are primarily
composed of Sn \emph{p}-like and S/Se/Te \emph{p}-like states,
respectively; (\emph{iv}) The parity of the top (bottom) of the valence
(conduction) band at $L$ is odd (even);
(\emph{v}) The band character around $L$ remains
unchanged upon including SOC.
These features already indicate that rock-salt SnS and SnSe are TCIs
just like SnTe~\cite{note1}. It should be emphasized here that, unlike
TIs~\cite{xia,zhang,chen2,lu,james,xiao2,sun,sun2}, the inverted band
order is found to be driven not by SOC but solely by chemical bonding
and crystal field [see Fig. \ref{band_inv} (j)].

The occurrence of band inversion at an even number of ${\bm k}$ points
(i.e., four equivalent $L$ points) and the fact that this band inversion is
not driven by SOC are suggestive of the formation of a crystal-symmetry
driven non-trivial topological state. In order to provide further support,
let us study the evolution of the band gap as a function of
the lattice constant. It is an obvious fact that any insulator is topologically
trivial in the atomic limit. Therefore, the occurrence of inverted band order
implies that the band gap has to close and re-open by progressively increasing
the lattice constant. This behavior is indeed found in Fig.~\ref{band_inv} (i):
The band gap $E_g$ at $L$ closes with increasing the lattice constant and then
re-opens with the opposite band character, i.e., the parity as well as the main
contributing weight of the constituent atoms being reversed for the top of the valence
band and the bottom of the conduction band [Fig.~\ref{band_inv} (d) and (h)].
These results clearly demonstrate that a topological phase
transition from a topologically non-trivial to a trivial states occurs
with increasing the lattice constant. Note that a similar behavior is
found even without including SOC [Fig.~\ref{band_inv}(c) and (g)],
indicating that SOC has indeed no influence in determining the band
character around $L$.
The evolution of the band character is schematically drawn in
Fig.~\ref{band_inv} (j).

\begin{figure} [htpb]
\begin{center}
\includegraphics[width=0.45\textwidth]{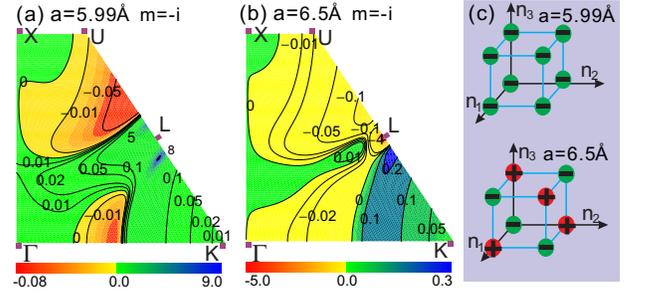}
\end{center}
\caption{(color online) Contour plots of the Berry curvature ${\bm\Omega}^m({\bm k})$ on the
($\overline{1}$10) mirror symmetry plane in the BZ for SnSe with the
experimental lattice constant $a$=5.99 \AA\,(a) and with an expanded
lattice constant $a$=6.5 \AA\,(b).
Only the component ${\Omega}^m_{\bot}({\bm k})$ perpendicular
to the ($\overline{1}$10) mirror plane with $m=-i$ is plotted. Note that
${\Omega}^{+i}_{\bot}({\bm k})=-{\Omega}^{-i}_{\bot}({\bm k})$.
The high symmetry momenta, ${\bm k}=n_1{\bm b}_1/2+n_2{\bm b}_2/2+n_3{\bm b}_3/2$, are
indicated in (a) and (b) by $\Gamma:\, (n_1,n_2,n_3)=(0,0,0)$, $X:\,(1,1,0)$,
$U:\, (\frac{5}{4},\frac{5}{4},\frac{1}{2})$, $L:\, (1,1,1)$, and
$K:\, (\frac{3}{4},\frac{3}{4},\frac{3}{2})$ with the reciprocal lattice vectors
${\bm b}_1={\frac{2\pi}{a}}(-1,1,1)$,
${\bm b}_2={\frac{2\pi}{a}}(1,-1,1)$, and ${\bm b}_3={\frac{2\pi}{a}}(1,1,-1)$.
(c)
The products of parity eigenvalues from the occupied states ($\delta_{n_1n_2n_3}=\pm1$)
at eight time reversal momenta ${\bm k}$, i.e., $n_1,n_2,n_3=0,1$, are indicated
for SnSe with both experimental and expanded lattice constants.
} \label{berry}
\end{figure}

To quantify the topological feature, we shall evaluate the mirror Chern number.
We first calculate the Berry curvature ${\bm\Omega}^m({\bm k})=\nabla_{\bm k}\times{\bm A^m({\bm k})}$ on the
($\overline{1}$10) mirror symmetry plane in the BZ. Here,
${\bm A^m({\bm k})}=i\sum_n\langle u_n^m(\bm k)|\nabla_{\bm k}|u_n^m(\bm k)\rangle$ is the Berry
connection, $u_n^m({\bm k})$ is the $n$-th eigenstate at momentum ${\bm k}$ and with
mirror eigenvalue $m\,(=\pm i)$ of the TB model described below,
and the sum is over all occupied bands. The results of the component ${\Omega}^m_{\bot}({\bm k})$
perpendicular to the ($\overline{1}$10) mirror plane are shown in Fig.~\ref{berry} for SnSe
with both experimental and expanded lattice constants. We find that the main contributions
are from momenta close to $L$ and that
${\Omega}^{+i}_{\bot}({\bm k})=-{\Omega}^{-i}_{\bot}({\bm k})$. We evaluate the mirror Chern
number~\cite{fu4}, $c_M=(n_{+i}-n_{-i})/2$, where $n_m=\int {\bm\Omega}^m({\bm k})\cdot d{\bm S}$,
and find that $c_M=-2\, (0)$ for SnSe with the experimental (expanded) lattice constant,
confirming the topological transition from a TCI to a trivial insulator with increasing the lattice constant.
We also calculate the $Z_2$ indices $(\nu_0; \nu_1\nu_2\nu_3)$~\cite{fu5} and find that this
index is $(0; 000)$
[see Fig.~\ref{berry} (c)] for SnSe with both experimental and expanded lattice constants, suggesting
that they are not time-reversal $Z_2$ TIs.

{\em Surface band structure.} Let us next examine the
intrinsic properties of the topological phase in SnS and SnSe, and
provide further evidence of this TCI state by inspecting the surface
properties. Unlike TIs for which an odd number of Dirac cones appears
in any surface orientation, TCIs have an even number of
topologically protected Dirac cones on high symmetry surfaces.
For the rock-salt crystal structure, gapless surface
states are expected to exist only on surfaces which are
perpendicular to the ($\overline{1}$10) mirror symmetry
planes~\cite{fu2,fu3,Safaeo}.

\begin{figure} [htpb]
\begin{center}
\includegraphics[width=0.45\textwidth]{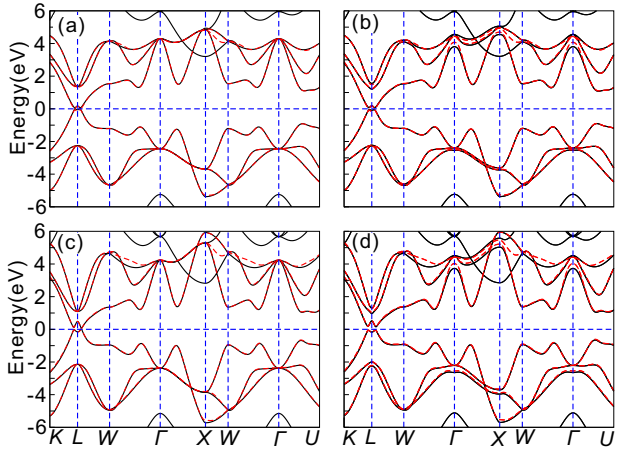}
\end{center}
\caption{(color online) Comparison of the bulk energy band
structures obtained from the {\it ab initio} TB model (red dashed
lines) and from the first-principles calculations (black solid
lines) for rock-salt SnS (a,b) and SnSe (c,d). Figures (a,c) and (b,d)
correspond to the cases without and with including SOC,
respectively.
} \label{band}
\end{figure}

To prove these expectations, we shall now compute the band dispersions for
the (001), (110) and (111) surfaces [see Fig.~\ref{crystal} (b)] using
the {\it ab initio} TB model. The {\it ab initio} TB model is
constructed by downfolding the bulk energy bands, obtained by
first-principles calculations, using maximally-localized Wannier
functions (MLWFs). As the bulk energy bands near Fermi energy
are predominantly formed by hybridized \emph{p}-like Sn and S/Se
orbitals, the MLWFs are derived from atomic \emph{p}-like orbitals
and the TB parameters are determined from the MLWFs overlap matrix.
The SOC is considered here in the atomic form:
\begin{align}
H_{\rm SO}^{p}(\lambda)=\frac{\lambda}{2}\left[\begin{array} {cccccc}
0 & 0 & -i & 0 & 0 & 1\\
0 & 0 & 0 & i & -1 & 0\\
i & 0 & 0 & 0 & 0 & -i\\
0 & -i & 0 & 0 & -i & 0\\
0 & -1 & 0 & i & 0 & 0\\
1 & 0 & i & 0 & 0 & 0
\end{array}\right]
\end{align}
with \emph{p}-like orbital bases
$\{|p_{x},\uparrow\rangle,|p_{x},\downarrow\rangle,|p_{y},\uparrow\rangle,|p_{y},\downarrow\rangle,|p_{z},\uparrow\rangle,|p_{z},\downarrow\rangle\}$,
where arrows indicate electron spins. The SOC parameter $\lambda$
for Sn, Se, and S are taken from experimental spectral data, i.e.,
$\lambda_{\rm Sn}$=0.27 eV, $\lambda_{\rm Se}$=0.22 eV, and
$\lambda_{\rm S}$=0.05 eV, respectively~\cite{wittel}. The quality
of the TB parametrization is successfully assessed in Fig.~\ref{band},
where the TB bulk band structures are compared with the
corresponding first-principles results.

\begin{figure}[htbp!]
\begin{center}
\includegraphics[width=0.45\textwidth]{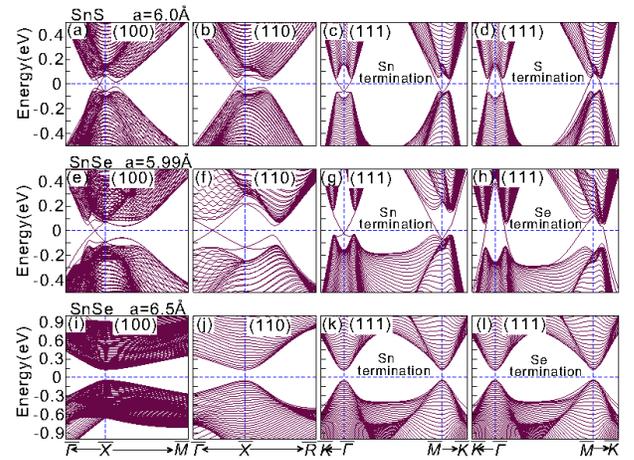}
\end{center}
\caption{(color online)
TB energy band structures of the (100),
(110), and (111) surfaces for rock-salt (a-d) SnS with
the experimental lattice constant $a$=6.0 \AA\,, (e-h) SnSe with the
experimental lattice constant $a$=5.99 \AA\,, and (i-l) SnSe with an
expanded lattice constant $a$=6.5 \AA\,. Notice that there exist two
distinct surface terminations for the (111) surface.
In (a)-(h) the
appearance of the gapless Dirac-cone-like metallic states evidences
the occurrence of TCI. With expanding the lattice constant in
(i)-(l), Dirac-cone-like metallic surface states disappear,
suggesting that the system is a trivial insulator.
The surface projected momenta are
indicated in Fig.~\ref{crystal}(b).
} \label{band_surface}
\end{figure}

Encouraged by this quantitative agreement, let us finaly compute the
surface band structures by adopting three slabs for
the (001), (110), and (111) surfaces with thickness of 89,
89, and 239 atomic layers, respectively.
The results of the TB calculations are summarized in Fig.~\ref{band_surface}. These results show clearly that
the (001), (110), and (111) surfaces %of both SnS and SnSe
with the experimental lattice constants
posses metallic states with opposite mirror eigenvalues which cross each other
forming a massless Dirac cone.
It have been shown~\cite{fu2,fu3} that %As already discussed in Refs.~\onlinecite{fu2} and \onlinecite{fu3},  for
the rock-salt TCIs with the mirror Chern number $c_M=-2$ guarantees the presence of two pairs of
counter-propagating, spin-resolved surface states with opposite mirror eigenvalues
along all symmetrically equivalent $\overline{\Gamma}$-${\overline X}$ lines in
the (100) surface, and only one pair in the (110) surface [Fig.~\ref{crystal} (b)].
Indeed, both SnS and SnSe surfaces follow this symmetry consideration,
displaying four equivalent Dirac cones in the (001) surface [Fig.~\ref{band_surface} (a) and (e)] and
two Dirac cones in the (110) surface [Fig. \ref{band_surface} (b) and (f)].
Instead, similarly to the case of SnTe~\cite{fu3}, four Dirac cones are found in the (111) surface,
one at $\overline{\Gamma}$ and other three at $\overline{M}$, as shown in
Fig.~\ref{band_surface} (c), (d), (g), and (h).

In addition, as already shown in Fig.~\ref{band_inv} (i) and Fig.~\ref{berry}, for the lattice constants larger than 6.10 \AA\, for SnS
and 6.37 \AA\, for SnSe, the band character changes and the systems undergo a topological transition
towards a trivial insulator. This is clearly reflected also
in the surface band structures computed for SnSe with the lattice constant $a=6.50$ \AA\, in
Fig.~\ref{band_surface} (i)-(l): the Dirac cones disappear and a broad spectral feature develops
on the top (bottom) of the valence (conduction) band with a finite band gap, a typical behavior of an
ordinary trivial insulator.
These results unambiguously demonstrates that rock-salt SnSe and SnS
represent the features of TCI.

In conclusion, using first-principles calculations together
with {\it ab initio} tight-binding model analyses, we have revealed
that rock-salt SnS
and SnSe represent a prime example of topological crystalline
insulators at ambient pressure without incorporating heavy elements.
We have shown that in both systems an even number of symmetry-protected Dirac cones
emerge in the (100), (110), and (111) surfaces perpendicular to the {($\overline{1}$10)} mirror
symmetry plane.
We have also shown that %although it is significantly reduced compared to SnTe,
the spin-orbit coupling is still important to open the band gap in the bulk phases although it is not necessary to
drive the topologically non-trivial state with the inverted band order, as proposed in the
original theory~\cite{fu}.
We have also demonstrated that a topological transition occurs toward a trivial insulator upon volume expansion.
Finally, we emphasize that the onset of the topological crystalline insulating
state in SnS and SnSe is
not dependent on alloying, strain, pressure, or any electronic structure engineering, but
SnS and SnSe are both topological crystalline insulators in their native phase.

This work was supported by the ``Hundred Talents Project'' of the
Chinese Academy of Sciences, NSFC of China (grand numbers 51074151
and 51174188), and Grant-in-Aid for Scientific Research from MEXT
Japan (grant numbers 24740269 and 25287096). We acknowledge Vienna
Scientific Cluster, Beijing Supercomputing Center of CAS
(including in Shenyang branch), and RIKEN Cluster of Clusters (RICC) facility
for computational resources.


\begin{thebibliography}{99}
\bibitem{note3} Although not shown for clarity, the same irreducible representations are assigned to the bands of SnS and SnTe.
Note that the origin of the symmetry operations is set to Sn atom.
\bibitem{note1} One of the important quantitative differences is that the band gap $E_g$ of SnSe ($E_g\sim0.202$ eV)
seems larger than SnTe ($E_g\sim0.172$ eV).
However, more careful study beyond GGA is required for better quantitative comparison.
\end{thebibliography}
\end{document}